\documentclass[12pt,draftcls,onecolumn,journal]{IEEEtran}
\usepackage{amsfonts}
\usepackage{amsfonts}
\usepackage{setspace}

\usepackage{enumerate}

\usepackage{amsmath}
\usepackage{amssymb}
\usepackage[dvips]{graphicx}
\usepackage{epsfig}
\usepackage{hyperref}

\newcommand{\tsnr}{{\text{\footnotesize{SNR}}}}
\newcommand{\tmin}{\text{min}}
\newcommand{\tmax}{\text{{max}}}
\newcommand{\E}{\mathbb{E}}

\newcommand{\Pb}{\bar{P}}

\newcommand{\figsize}{0.65}

\newcommand{\tMAC}{\text{MAC}}

\newcommand{\bR}{{\textbf{R}}}
\newcommand{\bD}{{\textbf{D}}}
\newcommand{\bS}{{\textbf{S}}}

\newcommand{\bbtheta}{\underline{\theta}}
\newcommand{\uutheta}{\breve{\theta}}
\newcommand{\vvtheta}{\overset{\circ}{\theta}}

\newtheorem{Lem1}{Proposition}%[section]
\newtheorem{Lem}{Theorem}
\newtheorem{Lemm}{Lemma}
\newtheorem{Rem}{Remark}

\newtheorem{Def}{Definition}

\begin{document}

\title{Effective Capacity of Buffer-Aided Full-Duplex Relay Systems with Selection Relaying}

%\author{\authorblockN{Michael Shell} \and
%\and \authorblockN{James Kirk\\ and Montgomery Scott}
%\authorblockA{Starfleet Academy\\
%San Francisco, California 96678-2391\\ Telephone: (800)
%555--1212\\ Fax: (888) 555--1212}}

% avoiding spaces at the end of the author lines is not a problem with
% conference papers because we don't use \thanks or \IEEEmembership

\author{\vspace{0.3cm}
\authorblockN{Deli Qiao}
\thanks{The author is with the School of Information Science and Technology, East China Normal University, Shanghai, China, 200241 (e-mail: dlqiao@ce.ecnu.edu.cn).}
\thanks{This work was supported in part by the National Natural Science Foundation of China under Grant 61172085. This paper will be presented in part at the IEEE Global Communications Conference (Globecom), San Diego, CA, Dec. 2015.}}

%\specialpapernotice{(Invited Paper)}

\maketitle

\begin{abstract}
In this work, the achievable rate of three-node relay systems with selection relaying under statistical delay constraints, imposed on the limitations of the maximum end-to-end delay violation probabilities, is investigated. It is assumed that there are queues of infinite size at both the source and relay node, and the source can select the relay or destination for data reception. Given selection relaying policy, the effective bandwidth of the arrival processes of the queue at the relay is derived. Then, the maximum constant arrival rate can be identified as the maximum effective capacity as a function of the statistical end-to-end queueing delay constraints, signal-to-noise ratios (\tsnr) at the source and relay, the fading distributions of the links, and the relay policy. Subsequently, a relay policy that incorporates the statistical delay constraints is proposed. It is shown that the proposed relay policy can achieve better performance than existing protocols. Moreover, it is demonstrated that buffering relay model can still help improve the throughput of relay systems in the presence of statistical delay constraints and source-destination link.
\end{abstract}

%\begin{spacing}{1.8}
\section{Introduction}

Relay channels can help improve the system coverage and throughput, and hence information-theoretic analysis of relay channels has been the research forefront for decades (see, e.g., \cite{relaycover}-\cite{selectionrelay}). For instance, Laneman \emph{et al.} in \cite{coopdiver} considered different relaying strategies, such as decode-and-forward (DF) and selection relaying, and showed that considerable cooperative diversity can be achieved with the relaying schemes. Among the relaying protocols, selection relaying schemes are attractive due to their potential to improve bandwidth utilization and cooperation diversity \cite{selectionrelay}. While providing powerful results, information-theoretic studies generally assume no buffer at the relay.

Recently, it has been shown that the achievable throughput can be further improved with the introduction of buffering relay model \cite{bufferrelay}. This is generally due to the information storage at the relay such that the shortcomings of existing relaying schemes caused by bad channel conditions can be overcome. The analysis of the buffer-aided relay systems has attracted much attention recently (see, e.g., \cite{schoberrelay}-\cite{directrelay} and references therein). For instance, the authors in \cite{schoberrelay} analyzed the two-hop relay system with buffer-aided relaying for adaptive and fixed rate transmission, and proposed
the throughput-optimal buffer-aided relaying protocols with significant performance improvements. In \cite{maxmaxrelay}, the authors proposed a max-max relay selection protocol, which chooses the source-relay and relay-destination link with the strongest channel gain. They found that this policy can achieve better performance than systems without buffering relay \cite{selectionrelay}. In \cite{directrelay}, the authors proposed a relay policy based on the relay link with strongest channel gain and the direct link that help improve the system performance. However, these works on buffer-aided relaying systems rarely consider the buffer at the source. In the presence of the buffer at the source, the delay experienced at the source buffer will be taken into account as well for the end-to-end delay, and the queue dynamics of the interacted queues are generally difficult to analyze. Note that in \cite{sdrrelay}, the authors investigated the buffer-aided relay systems with buffer at the source and energy-harvesting capability at each node, although the analysis is based on the average arrival and service rate, and only stability regions of different strategies are considered.

In this paper, we follow a different approach to analyze the buffer-aided relay systems. We consider the statistical delay constraints, imposed on the limitations of the maximum \emph{end-to-end} delay violation probabilities. We assume that there are buffers of infinite size at both the source and relay node, each subject to the statistical queueing constraints imposed on the limitations of buffer overflow probabilities. We consider the end-to-end delay composed of the queueing delay at the source and relay buffer. To handle the queueing dynamics of the relay networks, we employ the concept of effective bandwidth, which defines the bandwidth usage of given processes \cite{changbook}. More recently, Wu and Negi in \cite{dapeng} defined the dual concept of effective capacity, which
provides the maximum constant arrival rate that can be supported by
a given departure process while satisfying statistical delay
constraints. The analysis and application of effective capacity in various
settings have attracted much interest recently (see e.g. \cite{tangrelay}-\cite{deli-twohopend} and references therein). For instance, Tang and Zhang in \cite{tangrelay} analyzed the power allocation policies of relay networks with only one relay, where the relay node is assumed to have no buffer, i.e., the packets arriving to the relay node are forwarded immediately. In \cite{ecmultirelay}, Efazati and Azmi considered a multirelay network, and proposed a novel transmission scheme that selects different strategies such as best relay selection and distributed space-time coding based on the criterion that maximizes the obtained effective capacity. Still, there is no buffer at the relays in the considered system model. Parag and Chamberland in \cite{butterfly} provided a queueing analysis of a butterfly network with constant rate for each link, while assuming that there
is no congestion at the intermediate nodes. For the buffer-aided relay networks with buffer at the source, Du and Zhang in \cite{durelay} analyzed the throughput and power allocation policy in two-hop links under statistical end-to-end delay constraints, while imposing symmetric statistical delay constraints to the queues at the source and relay. The effective capacity of the two-hop link in the presence of statistical queueing constraints is given in \cite{deli-twohop}, and the performance of multi-relay links is analyzed in \cite{dualhop}. As a further step, we derived the maximum effective capacity of the two-hop links under statistical delay constraints with asymmetric delay constraints to the queues in \cite{deli-twohopend}. However, to the best knowledge of the author, there is no related work considering the buffer-aided relay systems with source-destination link.

In this work, we consider the buffer-aided relay systems with source-destination link. We assume that the channel state information (CSI) of all links is available at the source and relay, while the destination has no CSI of the source-relay link. We assume that the source employs selection relaying strategy, i.e., the source selects the relay or destination for data reception based on the CSI available. Since the CSI of the source-relay link is not available at the destination, we assume that the relay policy is informed to the destination via an one-bit acknowledge (ACK) signal such that the destination can perform successive decoding of the received signals when destination is selected for data reception. Given the relay protocol and statistical queueing constraints of the queues, we characterize the effective bandwidth of the arrival processes to the relay, which is one of the major findings of this work since it can be extended to various relay networks and constitutes an important basis for the ensuing effective capacity analysis. Then, based on the statistical delay tradeoff established in \cite{deli-twohopend}, we characterize the maximum effective capacity under the statistical delay constraints with fixed relay strategy. Also, we propose a relay scheme taking into account the statistical delay constraints. We show that the maximum effective capacity of the proposed variable relay schemes can be characterized similarly. Through numerical evaluation, this study reveals the benefits of buffering relay model in the presence of delay constraints and source-destination link. The main contributions of this paper are summarized as follows:
\begin{enumerate}
\item We provide a framework for analyzing the throughput of the relay systems with selection relaying under statistical delay constraints.
\item We determine the maximum effective capacity of the relay system with arbitrary selection relaying policy, and we obtain the limiting performance as the delay constraints vanish.
\item We propose a relay scheme based on statistical delay constraints that can further improve the achievable throughput compared with the existing best relay selection policy of buffer-aided relay systems.
\end{enumerate}

The rest of this paper is organized as follows. Section II discusses the necessary preliminaries on the system model, the statistical delay constraints and effective capacity.
In Section III, we present our main results on the effective capacity and relay policy, with numerical results given in Section IV. Finally, Section V concludes this paper, while some lengthy proofs are relegated to Appendix.

\section{Preliminaries}

\subsection{System Model}

\begin{figure}
\begin{center}
\includegraphics[width=\figsize\textwidth]{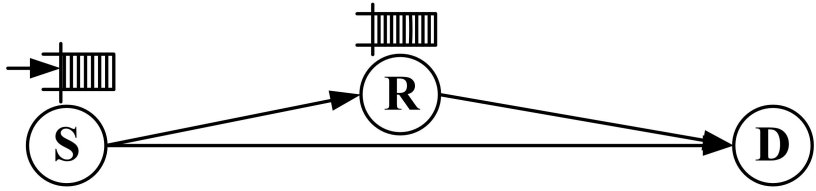}
\caption{The system model.}\label{fig:systemmodel}
\end{center}
\end{figure}

The three-node relay communication link is depicted in Fig.
\ref{fig:systemmodel}. The source can select the relay or destination for data reception. In this model, there are buffers of infinite size at both the source and relay node. In this work, we assume full-duplex relay such that transmission and reception can be performed simultaneously. Note that full-duplex relaying can be achieved through some form of analog self-interference cancellation followed by digital self-interference cancellation in the baseband domain \cite{fd-1}, \cite{fd-2}.% The relay just forwards the data from the source, i.e., it does not generate its own traffic.

The discrete-time input and output relationships in the $i$th symbol
duration are given by
\begin{align}
Y_r[i]&=g_{sr}[i]X_s[i]+I[i]+n_r[i] \\
Y_d[i]&=g_{sd}[i] X_s[i]+g_{rd}[i]X_r[i]+n_d[i]
\end{align}
where $X_j$ for $j\in\{s, r\}$ denote the input signal from the source $\bS$ and the relay $\bR$, respectively. The inputs are subject to individual average
energy constraints $\E\{|X_j|^2\}\le \Pb_j/B, j\in\{s,r\}$ where $B$
is the bandwidth. $Y_r, Y_d$ represent the received signal at the relay $\bR$ and the destination $\bD$, respectively. We assume that the fading coefficients $g_{sd}, g_{sr},  g_{rd}$ are jointly stationary and ergodic discrete-time
processes, and we denote the magnitude-square of the fading
coefficients by $z_{sd}[i]=|g_{sd}[i]|^2$ , $z_{sr}[i]=|g_{sr}[i]|^2$, and $z_{rd}[i]=|g_{rd}[i]|^2$. Note that $I[i]$ denotes the self-interference incurred by the full-duplex operation at the relay, which may include linear and nonlinear components of the transmitted signal $X_r[i]$ of the relay \cite{fd-2}. In \cite[Appendix A.1]{fd-1}, the received signal's SNR can be modeled as $\frac{\text{SNR}}{\gamma}$ with relative SNR loss $\gamma\ge1$ due to self-interference. Hence, we normalize the channel gain of the link $\bS-\bR$ by $\gamma$ to take into account the self-interference, i.e., $\tilde{z}_{sr}=z_{sr}/\gamma$. Denote $\mathbf{z} = (z_{sd},\tilde{z}_{sr},z_{rd})$. Assuming that there are $B$
complex symbols per second, we can easily see that the symbol energy
constraint of $\Pb_j/B$ implies that the channel input has a power
constraint of $\Pb_j$. Above, in the channel input-output
relationships, the noise component $n_j[i]$ is a zero-mean,
circularly symmetric, complex Gaussian random variable with variance
$\E\{|n_j[i]|^2\} = N_0$ for $j \in\{r,d\}$. The additive Gaussian noise
samples $\{n_j[i]\}$ are assumed to form an independent and
identically distributed (i.i.d.) sequence. We denote the
signal-to-noise ratio at source as $\tsnr=\frac{\Pb_s}{N_0 B}$, and at relay as $\tsnr_r = \frac{\Pb_r}{N_0 B}$.

\subsection{Statistical Delay Constraits}

%In this work, we seek to identify the maximum constant arrival rate to the source that can be supported by the relay system while satisfying the statistical delay constraints. Therefore, we need to guarantee that the data transmission through the relay, i.e., information flow over two queues, should satisfy the statistical delay constraints. %Then, the data transmission through direct link will satisfy the statistical delay constraints as well.
With the above mentioned settings, we first need the following result from \cite{changbook}.
\begin{Lem}[\cite{changbook}]
Suppose that the queue is stable and that both the arrival process
$a[n],n=1,2,\dots$ and service process $c[n], n=1,2,\dots$ satisfy
the G\"{a}rtner-Ellis limit, i.e., for all $\theta\ge0$, there
exists a differentiable logarithmic moment generating function
(LMGF) $\Lambda_A(\theta)$ such that\footnote{Throughout the text,
logarithm expressed without a base, i.e., $\log(\cdot)$, refers to
the natural logarithm $\log_e(\cdot)$.}
%\begin{small}
\begin{align}
\lim_{n\to\infty}\frac{\log \E\{e^{\theta\sum_{i=1}^n
a[n]}\}}{n}=\Lambda_A(\theta),
\end{align}
%\end{small}
and a differentiable LMGF $\Lambda_C(\theta)$ such that
%\begin{small}
\begin{align}\label{eq:lmgf}
\lim_{n\to\infty}\frac{\log \E\{e^{\theta\sum_{i=1}^n
c[n]}\}}{n}=\Lambda_C(\theta).
\end{align}
%\end{small}
If there exists a unique $\theta^*>0$ such that
\begin{align}
\Lambda_A(\theta^*)+\Lambda_C(-\theta^*)=0,
\end{align}
then
%\begin{small}
\begin{align} \label{eq:QoSexponentdef}
\lim_{Q_{\tmax}\to\infty}\frac{\log
\Pr\{Q>Q_{\tmax}\}}{Q_{\tmax}}=-\theta^*.
\end{align}
%\end{small}
where $Q$ is the stationary queue length.
\hfill $\blacksquare$
\end{Lem}
 This theorem tells us that the tail distribution of the stationary queue length decays exponentially. The proof of the theorem takes advantage of the large deviations principles (see e.g. \cite{ld-paper} and \cite{ld-book}) to characterize the exponential decay rate of buffer overflow probability. In particular, for large $Q_{\tmax}$, we have the approximation for the buffer violation probability: $\Pr\{Q>Q_{\tmax}\}\approx e^{-\theta^* Q_{\tmax}}$. Hence, while larger $\theta$ corresponds to more strict queueing constraints, smaller $\theta$ implies looser queueing constraints.

Assume that the first-in first-out (FIFO) queues are saturated, and hence they always attempt to transmit \cite{sdrrelay}. Then the queueing delay violation probability can be written equivalently as \cite{tangzhangcross2}, \cite{heathqos}
\begin{align}\label{eq:sddelay}
\Pr\{D>D_{\tmax}\} \doteq  e^{-J(\theta) D_\tmax}
\end{align}
where we defined $f(x)\doteq e^{cx}$ when $\lim_{x\to\infty}\frac{\log f(x)}{x}=c$, and
\begin{align}
J(\theta) = \theta \delta = -\Lambda_C(-\theta)
\end{align}
is the statistical delay exponent associated with the queue, with $\Lambda_C(\theta)$ the LMGF of the service rate, and $\delta$ is decided by the arrival and departure processes jointly. Now, we can express the probability density function of random variable $D$ as
\begin{align}
p_D(x) = \frac{\partial}{\partial x} \left(1-\Pr\{D>x\}\right) \doteq J(\theta) e^{-J(\theta) x}.
\end{align}

In this work, we seek to identify the maximum constant arrival rate to the source that can be supported by the relay system while satisfying the statistical delay constraints. Therefore, we need to guarantee that the data transmission of the flow with the largest \emph{end-to-end} delay should satisfy the statistical delay constraints, i.e., information flow over two queues. Consider two concatenated queues with statistical queueing constraints specified by $\theta_1$ and $\theta_2$, for queue 1 and queue 2, respectively. Given the queueing constraints specified by $\theta_1$ and $\theta_2$ with (\ref{eq:QoSexponentdef}) satisfied for each queue, we define
\begin{align}\label{eq:J1J2eq}
J_1(\theta_1)=-\Lambda_{C,1}(-\theta_1),\,\,\text{and}\,\, J_2(\theta_2)=-\Lambda_{C,2}(-\theta_2),
\end{align}
where $\Lambda_{C,1}(\theta_1)$ and $\Lambda_{C,2}(\theta_1)$ are the LMGF functions of the service rate of queue 1, 2, respectively. For data going through both queues, the end-to-end queueing delay violation probability can be characterized as\footnote{Note that the end-to-end delay consists of the queueing and transmission delays. As indicated in \cite[Section IV]{itcn}, the flow of data bits are treated as the flow of a fluid in the theory of effective bandwidth, in which case the transmission delay can be negligible if $T\ll D_\tmax$. The end-to-end delay can be approximated by the queueing end-to-end delay \cite{tangzhangcross2}, \cite{heathqos}.}
%\begin{small}
\begin{align}
&\Pr\{D_1+D_2>D_\tmax\}% \nonumber\\
%&
\doteq 1 - \int_0^{D_\tmax} \int_{0}^{D_\tmax - D_1}p_{D}(D_1)p_D(D_2)dD_2dD_1\nonumber\\
& = \left\{
\begin{array}{ll}
\frac{J_1(\theta_1)e^{-J_2(\theta_2)D_{\tmax}}-J_2(\theta_2)e^{-J_1(\theta_1)D_{\tmax}}}{J_1(\theta_1)-J_2(\theta_2)},& J_1(\theta_1)\neq J_2(\theta_2)\\
\left(1+J_1(\theta_1)D_{\tmax}\right)e^{-J_1(\theta_1)D_{\tmax}},&J_1(\theta_1)=J_2(\theta_2).
\end{array}\right.\label{eq:delayprob}
\end{align}
%\end{small}
Thereby, we need to guarantee that
\begin{align}\label{eq:queue12cond}
\Pr\{D_1+D_2>D_{\tmax}\}\le \varepsilon.
\end{align}
In this way, we can guarantee that the data transmissions through the relay, i.e., information flow over two queues, satisfy the statistical delay constraints. Then, the delay constraints of the whole system can be satisfied. Note that $(\varepsilon,D_\tmax)$ characterizes the statistical delay constraints with maximum delay violation probability $\varepsilon$ and maximum delay $D_\tmax$.

To facilitate the following analysis, we need the following tradeoff between $J_1(\theta_1)$ and $J_2(\theta_2)$.
\begin{Lemm}[\cite{deli-twohopend}]\label{lemm:J1J2relation}
Consider the following function
\begin{align}\label{eq:J1J2relation}
\vartheta(J_1(\theta_1),J_2(\theta_2))&=\frac{J_2(\theta_2)e^{-J_1(\theta_1)D_\tmax} - J_1(\theta_1)e^{-J_2(\theta_2)D_\tmax}}{J_2(\theta_2)-J_1(\theta_1)} \nonumber\\
&= e^{-J_0 D_\tmax}=\varepsilon, \,\text{for} \, 0\le\varepsilon\le1,
\end{align}
where $J_0=-\frac{\log(\varepsilon)}{D_\tmax}$ is defined as the statistical delay exponent associated with $(\varepsilon,D_\tmax)$. Denoting $J_2(\theta_2) = \Phi(J_1(\theta_1))$ as a function of $J_1(\theta_1)$, we have
\begin{enumerate}[a)]

\item $\Phi$ is continuous. For $J_1(\theta_1)=J_{th}(\varepsilon)$, we have
\begin{align}
\Phi(J_1(\theta_1)) =  J_{th}(\varepsilon)
\end{align}
where
\begin{align}\label{eq:Jfunctioncond}
J_{th}(\varepsilon) = -\frac{1}{D_{\tmax}}\left(1+\mathcal{W}_{-1}\left(-\frac{\varepsilon}{e}\right)\right)
\end{align}
where $\mathcal{W}_{-1}(\cdot)$ is the Lambert W function, which is the inverse function of $y=xe^x$ in the range $(-\infty,-1]$.

\item $\Phi$ is strictly decreasing in $J_1(\theta_1)$.

\item $\Phi$ is convex in $J_1(\theta_1)$.

\item $J_1(\theta_1)\in[J_0,\infty)$, and $J_2(\theta_2)=\Phi(J_1(\theta_1))\in[J_0,\infty)$.

\end{enumerate}
\end{Lemm}

%\begin{Rem}
This lemma indicates that to have the end-to-end delay constraints satisfied, we must increase $J_1(\theta_1)$ if $J_2(\theta_2)$ is decreased; vice versa.
%\end{Rem}
%\begin{figure}
%\begin{center}
%\includegraphics[width=\figsize\textwidth]{delaybound=1e-3_dmax=1.eps}
%\caption{$J_2$ v.s. $J_1$. $D_\tmax=1$. $\varepsilon=0.001$.}\label{fig:delaybound=1e-3_dmax=1}
%\end{center}
%\end{figure}

\subsection{Effective Capacity}\label{sec:ecdef}

Denote the queue at source $\bS$ as queue 1, and the queue at relay $\bR$ as queue 2. Denote $\Omega$ as the set of pairs $(\theta_1,\theta_2)$ such that (\ref{eq:queue12cond}) can be satisfied. Assume $\theta_1>0$ and $\theta_2>0$ at the source and the relay node with $(\theta_1,\theta_2)\in\Omega$. %\footnote{Note that the optimal values of $\theta_1$ and $\theta_2$ are determined through iterating over the lower boundary curve specified by Lemma \ref{lemm:J1J2relation} to achieve the maximum effective capacity as specified in Theorem \ref{theo:ecresultfix} later. }.
Assume that the constant arrival rate at the source is $R\ge0$, and the channels operate at their capacities. Then, the effective capacity with statistical queueing constraints $(\theta_1,\theta_2)$ is defined as the maximum constant arrival rate such that both the queueing constraints can be satisfied. More specifically, to satisfy the queueing constraint at the source, we should have
\begin{align}\label{eq:theta1cond}
\tilde{\theta}\ge\theta_1
\end{align}
where $\tilde{\theta}$ is the solution to
\begin{align}\label{eq:queue1cond}
R = -\frac{\Lambda_{C,1}(-\tilde{\theta})}{\tilde{\theta}}
\end{align}
and $\Lambda_{C,1}(\theta)$ is the LGMF of the service rate for queue 1, i.e., queue at the source.

Also, in order to satisfy the queueing constraint of the relay node $\bR$, we must have
\begin{align}\label{eq:theta2cond}
\hat{\theta}\ge\theta_2
\end{align}
where $\hat{\theta}$ is the solution to
\begin{align}\label{eq:queue2cond}
\Lambda_{A,2}(\theta)+\Lambda_{C,2}(-\theta)=0.
\end{align}
where $\Lambda_{A,2}(\theta)$ is the LGMF of the arrival process to queue 2, $\Lambda_{C,2}(\theta)$ is the LGMF of the service process of queue 2, i.e., queue at the relay.

Note that we can derive the effective capacity $R_E(\theta_1,\theta_2)$ with $(\theta_1,\theta_2)$ following the method provided in \cite[Theorem 2]{deli-twohop}. After these characterizations, effective capacity of the buffer-aided relay system under statistical delay constraints $(\varepsilon,D_\tmax)$ can be formulated as follows.
\begin{Def}\label{def:ecdef}
The effective capacity of the buffer-aided relay system with statistical delay constraints specified by $(\varepsilon,D_\tmax)$ is given by
\begin{align}\label{eq:effdefi}
R_\varepsilon(\varepsilon,D_\tmax)=\sup_{(\theta_1,\theta_2)\in\Omega} R_E(\theta_1,\theta_2).
\end{align}
Hence, effective capacity is now the maximum constant arrival rate that can be supported by the relay system under statistical delay constraints.
\end{Def}

\section{Effective Capacity with Selection Relaying Protocol in Block-Fading Channel}

In the following, we first discuss the transmission strategy in detail to obtain the associated channel rate of the links. Then, assuming that $(\theta_1,\theta_2)$ are given, we obtain the effective bandwidth of the arrival processes of the queue at the relay given selection relaying policy. Next, we derive the effective capacity for arbitrary relay policy in a general form. Afterwards, we propose a relay policy taking into account the statistical delay constraints.

\subsection{Transmission Strategy and Channel Rate}

We assume that perfect CSI of all links is available at $\bS$ and $\bR$, while only the CSI
of the links $\bS-\bD$ and $\bR-\bD$ is available at $\bD$. The transmission power levels
at the source and relay are fixed and hence no
power control is employed (i.e., nodes are subject to short-term power constraints).
We further assume that the channel capacity for each link can be achieved, i.e., the service processes
are equal to the instantaneous Shannon capacities of the links. We
consider a block fading scenario in which the fading stays constant
for a block of $T$ seconds and changes independently from one block
to another.

\begin{figure}
\begin{center}
\includegraphics[width=\figsize\textwidth]{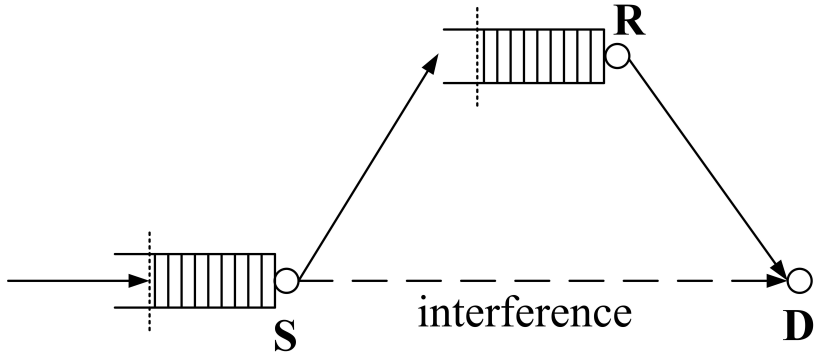}
\caption{Select relay for sending information.}\label{fig:srphase}
\end{center}
\end{figure}

We consider selection relaying protocols, i.e., the source selects the relay for data reception when the channel condition at the relay is larger than certain threshold \cite{coopdiver}. Denote $\mathcal{Z}$ as the region such that when $\mathbf{z}\in\mathcal{Z}$, the source $\bS$ selects the relay $\bR$ for data reception. Therefore, $\mathcal{Z}$ stands for the relay strategy employed in the system.
The relay strategy is forwarded by the source to the destination through an one-bit acknowledge (ACK) signal such that the destination can perform successive decoding of the received signals when $\mathbf{z}\in\mathcal{Z}^c$.
%In the following, we assume $\mathcal{Z}\neq \emptyset$, since when $\mathcal{Z}=\emptyset$, the channel reduces to the point-to-point link.

\begin{figure}
\begin{center}
\includegraphics[width=\figsize\textwidth]{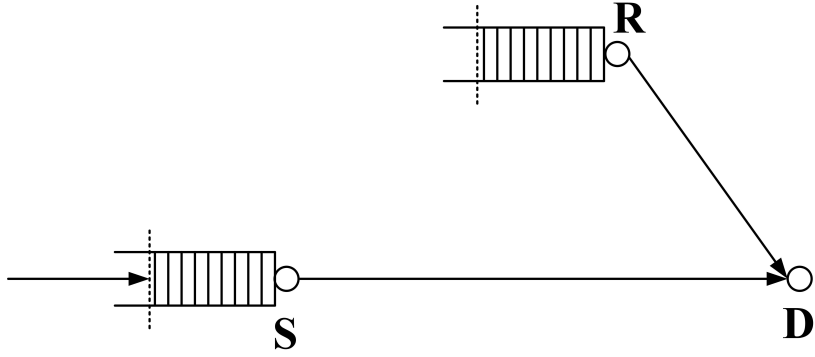}
\caption{Select destination for sending information.}\label{fig:sdphase}
\end{center}
\end{figure}

When $\mathbf{z}\in\mathcal{Z}$, due to the buffer at the relay, the transmitted messages from the source and the relay node are different. So we cannot exploit spatial diversity as we can when there is no buffer at the relay. In this case, we have a two-hop channel while the transmitted signal of $\bS-\bR$ link forms interference to the $\bR-\bD$ link. See Fig. \ref{fig:srphase} for the illustration of the information flow. Then the instantaneous service rates for the queues at the source and relay node are given by
\begin{align}
C_{sr} &= TB\log_2(1+\tsnr \tilde{z}_{sr}),\\
C_{rd} & = TB\log_2\left(1+\frac{\tsnr_r z_{rd}}{1+\tsnr z_{sd}}\right).
\end{align}
When $\mathbf{z}\in\mathcal{Z}^c$, $\bS$ selects the destination $\bD$ for data reception. In this case,
we have a two-user multiple access channel. See Fig. \ref{fig:sdphase} for the illustration of the information flow.
Note that the destination can perform successive decoding of the received signal from the source and relay node. Define $\mathcal{Z}_{0}$ as the region depending on $z_{sd}$ and $z_{rd}$ such that when $\mathbf{z}\in\mathcal{Z}^c\cap\mathcal{Z}_0$, the destination $\bD$ decodes the received signal in the order of $(\bR,\bS)$, i.e., the sent signal from the source sees no interference, and the instantaneous rates are given by \cite{Cover}
\begin{align}
C_{sd} &= TB\log_2(1+\tsnr z_{sd}),\\
C_{rd} & = TB\log_2\left(1+\frac{\tsnr_r z_{rd}}{1+\tsnr z_{sd}}\right).
\end{align}

On the other hand, when $\mathbf{z}\in\mathcal{Z}^c\cap\mathcal{Z}_0^c$, the decoding order at $\bD$ is $(\bS,\bR)$, i.e., the sent signal from the relay sees no interference, and the instantaneous rates are given by
\begin{align}
C_{sd} &= TB\log_2\left(1+\frac{\tsnr z_{sd}}{1+\tsnr_r z_{rd}}\right),\\
C_{rd} & = TB\log_2\left(1+\tsnr_r z_{rd}\right).
\end{align}

To summarize, we have the service rates of the queues at the source and relay node as
%\begin{small}
\begin{align}\label{eq:rates}
C_{s}=\left\{
\begin{array}{ll}
TB\log_2\left(1+\tsnr \tilde{z}_{sr}\right),&\mathbf{z}\in \mathcal{Z}\\
TB\log_2(1+\tsnr z_{sd}),&\mathbf{z}\in\mathcal{Z}^c\cap \mathcal{Z}_0\\
TB\log_2\left(1+\frac{\tsnr z_{sd}}{1+\tsnr_r z_{rd}}\right), &\mathbf{z}\in\mathcal{Z}^c\cap \mathcal{Z}_0^c
\end{array}
\right.
\end{align}
%\end{small}
and
%\begin{small}
\begin{align}\label{eq:rater}
%\hspace{-.5cm}
C_{r}=\left\{
\begin{array}{ll}
TB\log_2\left(1+\frac{\tsnr_r z_{rd}}{1+\tsnr z_{sd}}\right),&\mathbf{z}\in \mathcal{Z}\cup \{\mathcal{Z}^c\cap \mathcal{Z}_0\}\\
TB\log_2\left(1+\tsnr_r z_{rd}\right), &\mathbf{z}\in\mathcal{Z}^c\cap \mathcal{Z}_0^c
\end{array}\right.
\end{align}
%\end{small}
respectively. Above, the rates are in the units of bits per block or equivalently bits per $T$ seconds. These can be regarded as the service processes of the queues at the source and relay, respectively.

To ensure the stability of the queue at the relay, we need to enforce the following condition \cite{changbook}
\begin{align}
\E_{\mathbf{z}\in \mathcal{Z}}\{TB\log_2\left(1+\tsnr \tilde{z}_{sr}\right)\} < \E_{\mathbf{z}}\{C_{r}\}\label{eq:bufferstab2}
\end{align}
where $\E_{\mathbf{z}\in\mathcal{Z}}\{\cdot\}$ denotes the expectation over the region $\mathcal{Z}$, i.e., $\int_{\mathbf{z}\in\mathcal{Z}}\left\{\cdot\right\}p_{\mathbf{z}}(\mathbf{z})d\mathbf{z}$ with $p_{\mathbf{z}}(\mathbf{z})$ the probability density function of $\mathbf{z}$. The above equation implies that the average arrival rate of the queue at the relay should be less than the average service rate. We assume that the above condition is satisfied for the relay policies in consideration, otherwise the effective capacity is deemed as zero.

\subsection{Effective Bandwidth of the $\bS$-$\bR$ Link}

Note that $\Lambda_{C,1}(\theta)$ and $\Lambda_{C,2}(\theta)$ can be easily derived with the relay policy $\mathcal{Z}$ after we obtain the service rate of queue 1 and 2 in (\ref{eq:rates}) and (\ref{eq:rater}), respectively. Then, from the discussions in Section \ref{sec:ecdef}, we must obtain the effective bandwidth of the arrival processes of the queue at the relay to derive the effective capacity.

This can be achieved by borrowing the idea of intree-network \cite{changbook}. Note that queue 1 is the \emph{predecessor} of queue 2, or equivalently, queue 2 is the \emph{successor} of queue 1. Now the three-node relay network can be viewed as an \emph{intree network}. The routing depends on the relay protocol designed. Following the definitions in \cite[Section 9.4]{changbook}, we define the routing variable
\begin{align}
p_1[i] =\left\{\begin{array}{ll}
1,& \mathbf{z}\in\mathcal{Z},\\
0,&\mathbf{z}\in\mathcal{Z}^c.
\end{array}\right.
\end{align}
That is, $p_1[i]= 1$ if the departure from queue 1 at $i$th symbol is routed to the successor, i.e., queue 2, and $p_1[i]=0$ if the departure from queue 1 at $i$th symbol leaves the intree network, i.e., goes to the destination node. Now, we have the log-moment generating function for the routing process as
%\begin{small}
\begin{align}\label{eq:lmgf-routing}
\Lambda_{p_1}(\theta,\mathcal{Z}) &= \lim_{n\to\infty}\frac{\log\E\left\{e^{\theta \sum_{i=1}^np_1[i]}\right\}}{n}.
\end{align}
%\end{small}

Recalling the definitions in Section \ref{sec:ecdef}, we have the following result.
\begin{Lem1}
Given $(\theta_1,\theta_2)$ and the routing process specified by $\mathcal{Z}$, the LMGF of the departure process from the source to the relay, or equivalently the arrival process to the relay node, is given by
%\begin{small}
\begin{align}\label{eq:ebsrdef}
\Lambda_{A,2}(\theta) =\left\{
\begin{array}{ll}
 R\Lambda_{p_1}(\theta,\mathcal{Z}),& 0\le\Lambda_{p_1}(\theta,\mathcal{Z})\le \tilde{\theta},\\
 R\tilde{\theta} + \Lambda_{C,1}\left(\Lambda_{p_1}(\theta,\mathcal{Z})-\tilde{\theta}\right),&\Lambda_{p_1}(\theta,\mathcal{Z})> \tilde{\theta}.
\end{array}
\right.
\end{align}
%\end{small}
where $R$ is the constant arrival rate to queue 1, and $\tilde{\theta}$ is defined in (\ref{eq:queue1cond}).
\end{Lem1}
\emph{Proof: }Note that the only arrival to the relay node is from the source. Following the procedures described in \cite[Section 9.4]{changbook}, we know that
\begin{align}
\Lambda_{A,2}(\theta) = \Xi_1(\Lambda_{p_1}(\theta,\mathcal{Z}))
\end{align}
where $\Xi_1(\cdot)$ is the effective bandwidth of the departure process from queue 1. Since the arrival to the source is constant, we have
\begin{align}
\Xi_1(\theta) = \left\{\begin{array}{ll}
\theta R, & 0\le\theta\le \tilde{\theta}\\
\tilde{\theta}R + \Lambda_{C,1}(\theta-\tilde{\theta}),& \theta>\tilde{\theta}.
\end{array}
\right.
\end{align}
Substituting the value of $\Lambda_{p_1}(\theta,\mathcal{Z})$ to the above equation yields the result directly.\hfill$\square$

\subsection{Effective Capacity}

Under the block fading assumption, the logarithmic moment generating
functions for the service processes of queues at the source $\bS$ and the relay $\bR$ as
functions of $\theta$ are given by \cite{tangrelay}
\begin{align}\label{eq:preequ-1}
\hspace{-.5cm}\Lambda_{C,1}(\theta)&=\log \E\left\{e^{\theta C_s}\right\} \quad\text{and}\quad
\Lambda_{C,2}(\theta)=\log \E\left\{e^{\theta C_{r}}\right\}
\end{align}
where $C_{s}$ and $C_{r}$ are given by (\ref{eq:rates}) and (\ref{eq:rater}), respectively. Now, due to the assumption that the fading changes independently from one block to another, we can, for instance, simplify the logarithmic moment generating function in (\ref{eq:lmgf}) as $\Lambda_C = \lim_{n\to\infty}\frac{\log \E\{e^{\theta\sum_{i=1}^n
c[i]}\}}{n} = \lim_{n\to\infty}\frac{\log \prod_{i=1}^n \E\{e^{\theta c[i]}\}}{n} = \lim_{n\to\infty}\frac{\sum_{i=1}^n\log\E\{e^{\theta c[i]}\}}{n} = \lim_{n\to\infty}\frac{n\log\E\{e^{\theta c[1]}\}}{n} = \log\E\{e^{\theta c[1]}\}$. If fading is correlated, such simplifications are in general not possible and analysis needs to be based on the limit forms of the asymptotic logarithmic moment generating functions. However, if the service rates can be regarded as Markov modulated processes, then it is shown in \cite[Section 7.2]{changbook} that $\lim_{n\to\infty}\frac{\log \E\{e^{\theta\sum_{i=1}^n
c[i]}\}}{\theta n} = \frac{1}{\theta} \log \text{sp} (\phi(\theta)r)$
where $\text{sp}(A)$ denotes the spectral radius or equivalently the maximum of the absolute values of the eigenvalues of the matrix $A$, and $\phi(\theta)r$ is a matrix which depends on the transition probabilities of the Markov process. Similarly, we can derive the limit forms of the asymptotic logarithmic moment generating function of $p_1[i]$, $\lim_{n\to\infty}\frac{\log \E\{e^{\theta\sum_{i=1}^n
p_1[i]}\}}{n} =  \log \text{sp} (\varphi(\theta))$, if the routing process can be viewed as Markov modulated processes with $\varphi(\theta)$ representing the transition matrix of the associated Markov process. In such cases, an analysis similar to the one given in this paper can be pursued to identify the maximal effective capacity.

Combining (\ref{eq:preequ-1}) with (\ref{eq:J1J2eq}) gives us $J_1(\theta_1)$ and $J_2(\theta_2)$ as follows
\begin{align}
J_1(\theta_1)&=-\log\Bigg(\E_{\mathbf{z}\in\mathcal{Z}}\left\{e^{-\theta_1TB\log_2(1+\tsnr \tilde{z}_{sr})}\right\} + \E_{\mathbf{z}\in\mathcal{Z}^c\cap\mathcal{Z}_0}\left\{e^{-\theta_1 TB\log_2(1+\tsnr z_{sd})}\right\} \nonumber\\
&\hspace{2cm}+ \E_{\mathbf{z}\in\mathcal{Z}^c\cap\mathcal{Z}_0^c}\left\{e^{-\theta_1 TB\log_2\left(1+\frac{\tsnr z_{sd}}{1+\tsnr_r z_{rd}}\right)}\right\}\Bigg),\label{eq:bfJ1}\\
J_2(\theta_2)&=-\log\Bigg(\E_{\mathbf{z}\in\mathcal{Z}\cup(\mathcal{Z}^c\cap\mathcal{Z}_0)}\left\{e^{-\theta_2TB\log_2\left(1+ \frac{\tsnr_r z_{rd}}{1+\tsnr z_{sd}}\right)}\right\}  + \E_{\mathbf{z}\in\mathcal{Z}^c\cap\mathcal{Z}_0^c}\left\{e^{-\theta_2 TB\log_2(1+\tsnr_r z_{rd})}\right\}\Bigg).\label{eq:bfJ2}
\end{align}

Now, the LMGF for the routing process (\ref{eq:lmgf-routing}) can be written as
%\begin{small}
\begin{align}\label{eq:lmgf-routing-bf}
\Lambda_{p_1}(\theta,\mathcal{Z}) & = \log\left(\Pr\{\mathbf{z}\in\mathcal{Z}^c\} + e^{\theta}\Pr\{\mathbf{z}\in\mathcal{Z}\} \right).
\end{align}
%\end{small}
According to (\ref{eq:ebsrdef}), the LMGF for the arrival process of the queue at the relay is
%\begin{small}
\begin{align}\label{eq:relayareb}
\Lambda_{A,2}(\theta) =\left\{
\begin{array}{ll}
 R\Lambda_{p_1}(\theta,\mathcal{Z}),& 0\le\Lambda_{p_1}(\theta,\mathcal{Z})\le \tilde{\theta},\\
 R\tilde{\theta} + \log\E\left\{e^{ \left( \Lambda_{p_1}(\theta,\mathcal{Z}) - \tilde{\theta}\right) C_s}\right\},&\Lambda_{p_1}(\theta,\mathcal{Z})> \tilde{\theta}.
\end{array}
\right.
\end{align}
%\end{small}

For the following analysis, we need to characterize the relationship between $\Lambda_{p_1}(\theta,\mathcal{Z})$ and $\theta$. We have the following result.
\begin{Lemm}\label{lemm:lambdap}
Consider the function
\begin{align}\label{eq:routingdef}
\Lambda_{p_1}(\theta,\mathcal{Z}) = \log\left(\Pr\{\mathbf{z}\in\mathcal{Z}^c\}+ e^{\theta}\Pr\{\mathbf{z}\in\mathcal{Z}\}\right).
\end{align}
This function has the following properties:
\begin{enumerate}[a)]
\item $\Lambda_{p_1}(0,\mathcal{Z}) = 0$, and $\Lambda_{p_1}(\theta,\mathcal{Z})\le\theta$.

\item $\Lambda_{p_1}(\theta,\mathcal{Z})$ is increasing in $\theta$.

\item $\Lambda_{p_1}(\theta,\mathcal{Z})$ is a convex function of $\theta$.
\end{enumerate}
\end{Lemm}
\emph{Proof:} \begin{enumerate}[a)]
\item This property can be readily seen by evaluating the function at $\theta = 0$, and noting that $\Pr\{\mathbf{z}\in\mathcal{Z}\}\le1$.

\item The first derivative of $\Lambda_{p_1}$ with respect to $\theta$ can be evaluated as
\begin{align}\label{eq:lambdathetader}
\dot{\Lambda}_{p_1}(\theta)=\frac{e^{\theta}\Pr\{\mathbf{z}\in\mathcal{Z}\}}{\Pr\{\mathbf{z}\in\mathcal{Z}^c\}+ e^{\theta}\Pr\{\mathbf{z}\in\mathcal{Z}\}}>0.
\end{align}
Hence, it is increasing in $\theta$.

\item This property follows immediately since $e^{\theta}$ is log-convex, and non-negative multiplication does not alter the convexity.
    \end{enumerate}
    \hfill$\square$

\begin{Rem}
Therefore, when $\theta=0$, we have $\Lambda_{p_1}(\theta,\mathcal{Z})=0$. As $\theta$ increases, $\Lambda_{p_1}(\theta,\mathcal{Z})$ increases. Also, $\Lambda_{p_1}(\theta,\mathcal{Z})$ increases at least linearly with $\theta$. Note that $\frac{\theta}{\Lambda_{p_1}(\theta,\mathcal{Z})}$ is decreasing in $\theta$, since the $\frac{\Lambda_{p_1}(\theta,\mathcal{Z})}{\theta}$ is increasing in $\theta$ \cite{changbook}. Considering (\ref{eq:lambdathetader}), we have
\begin{align}
\lim_{\theta\to0}\frac{\theta}{\Lambda_{p_1}(\theta,\mathcal{Z})} = \frac{1}{\dot{\Lambda}_{p_1}(0)}= \frac{1}{\Pr\{\mathbf{z}\in\mathcal{Z}\}}
\end{align}
where L'Hospital's rule is used.
\end{Rem}

From Lemma \ref{lemm:lambdap}, we can show that $\frac{J_2(\theta_2)}{\Lambda_{p_1}(\theta_2,\mathcal{Z})}$ is still a decreasing function of $\theta_2$ similar to \cite[Lemma 1]{deli-twohop}, which is fundamental to this article.

With the selection relaying strategy $\mathcal{Z}$, we can establish an upperbound on the arrival rates supported by the relay system with any specific $(\theta_1,\theta_2)$.
\begin{Lem1}\label{prop:upperbound}
The constant arrival rates, which can be supported by the buffering relay system with statistical queueing constraints specified by $(\theta_1,\theta_2)$ at the source and relay, respectively, are upperbounded by
%\begin{small}
\begin{align}\label{eq:upperboundrate}
R&\le\min\bigg\{-\frac{1}{\theta_1}\log\E\left\{e^{-\theta_1C_s}\right\}, %\nonumber\\ &\hspace{.5cm}
-\frac{1}{\Lambda_{p_1}(\theta_2,\mathcal{Z})}\log\E\left\{e^{-\theta_2C_{r}}\right\}\bigg\}\\
&=\min\left\{\frac{J_1(\theta_1)}{\theta_1}, \frac{J_2(\theta_2)}{\Lambda_{p_1}(\theta_2,\mathcal{Z})}\right\}
\end{align}
%\end{small}
\end{Lem1}
\emph{Proof:} See Appendix \ref{app:upperbound}. \hfill$\square$
%\begin{Rem}
%The above proposition is necessary for the following derivation. It can be understood easily,
%\end{Rem}

Define
%\begin{small}
\begin{align*}%\label{eq:omegaeps}
\Omega_{\varepsilon} = \{(\theta_1,\theta_2): \text{ $J_1(\theta_1)$ and $J_2(\theta_2)$ are solutions to }\,(\ref{eq:J1J2relation})\}.
\end{align*}
%\end{small}
Similar to the discussions in \cite{deli-twohopend} regarding two-hop channels, we can iterate over $(J_1(\theta_1),J_2(\theta_2))$ satisfying (\ref{eq:J1J2relation}) and obtain the following result. Note that $z_{ij,\tmax}$ and $z_{ij,\tmin}$ represent the largest or smallest channel gain of link $i-j$, respectively.
\begin{Lem}\label{theo:ecresultfix}
The effective capacity of the buffer-aided relay systems with selection relaying strategy $\mathcal{Z}$  subject to statistical delay constraints specified by $(\varepsilon,D_{\tmax})$ is given by the following:

\vspace{.3cm}
\textbf{\underline{Case I}}: If $\theta_{1,th}= \Lambda_{p_1}(\theta_{2,th},\mathcal{Z})$,
%\begin{small}
\begin{gather}
\hspace{-.5cm}R_\varepsilon(\varepsilon,D_\tmax)=\frac{J_{th}(\varepsilon)}{\theta_{1,th}},
\end{gather}
%\end{small}
where ($\theta_{1,th}$,$\theta_{2,th}$) is the unique solution pair to $J_1(\theta_1)=J_{th}(\varepsilon)$, and $J_2(\theta_2)=J_{th}(\varepsilon)$.

\textbf{\underline{Case II}}: If $\theta_{1,th}>\Lambda_{p_1}(\theta_{2,th},\mathcal{Z})$,
%\begin{small}
\begin{gather}\label{eq:ecresultcase2}
\hspace{-.5cm}R_\varepsilon(\varepsilon,D_\tmax)= \left\{
\begin{array}{ll}
\frac{J_0}{\theta_{1,0}},& TB\log_2\left(1+\frac{\tsnr_r z_{rd,\tmin}}{1+\tsnr z_{sd,\tmax}}\right),\\
& \ge \max\{TB\log_2(1+\tsnr \tilde{z}_{sr,\tmax}),TB\log_2(1+\tsnr z_{sd,\tmax})\}\\
\frac{J_{1}(\vvtheta_1)}{\vvtheta_{1}},&\text{otherwise.}
\end{array}
\right.
\end{gather}
%\end{small}
where $\theta_{1,0}$ is the solution to $J_1(\theta_1) = J_0$, and $\vvtheta_1$ is the smallest value of $\theta_1$ with $(\theta_1,\theta_2)\in\Omega_\varepsilon$ satisfying
\begin{align}\label{eq:fixresultcond}
&-\frac{1}{\theta_1}\log\E\left\{e^{-\theta_1 C_s}\right\}%\nonumber\\
%&
=-\frac{1}{\theta_1}\Big(\log\E\left\{e^{-\theta_2
C_{r}}\right\}\nonumber\\
&\hspace{1.5cm}
+\log\E\left\{e^{(\Lambda_{p_1}(\theta_{2},\mathcal{Z})-\theta_1)
C_s}\right\}\Big).
\end{align}
Moreover, if $\frac{d J_2(\theta)}{d\theta}\big|_{\theta=\bbtheta_1}\le \frac{d J_1(\theta)}{d\theta}\big|_{\theta=\bbtheta_1}$, where $\bbtheta_1$ is given by $(\theta_1,\theta_2)\in\Omega_\varepsilon$ with
\begin{align}
\theta_1=\Lambda_{p_1}(\theta_{2},\mathcal{Z}),
\end{align}
then the solution to (\ref{eq:fixresultcond}) with $(\theta_1,\theta_2)\in\Omega_\varepsilon$ is unique.

\textbf{\underline{Case III}}: If $\theta_{1,th}<\Lambda_{p_1}(\theta_{2,th},\mathcal{Z})$,
%\begin{small}
\begin{gather}\label{eq:ecresultcase3}
\hspace{-.3cm}R_\varepsilon(\varepsilon,D_\tmax)=
\left\{
\begin{array}{ll}
\frac{J_0}{\Lambda_{p_1}(\theta_{2,0},\mathcal{Z})},&\min\Big\{TB\log_2(1+\tsnr \tilde{z}_{sr,\tmin}),\\
&TB\log_2\left(1+\frac{\tsnr z_{sd,\tmin}}{1+\tsnr_r z_{rd,\tmax}}\right)\Big\}\\
&\ge\frac{J_0}{\Lambda_{p_1}(\theta_{2,0},\mathcal{Z})}\\
\frac{J_{2}(\uutheta_2)}{\Lambda_{p_1}(\uutheta_2,\mathcal{Z})},&\text{otherwise.}
\end{array}
\right.
\end{gather}
%\end{small}
where $\theta_{2,0}$ is the solution to $J_2(\theta_2) = J_0$, and ($\uutheta_1$,$\uutheta_2$) is the unique solution to
\begin{align}\label{eq:fixresultcond2}
\frac{J_{1}(\theta_1)}{\theta_1} = \frac{J_{2}(\theta_2)}{\Lambda_{p_1}(\theta_2,\mathcal{Z})}
\end{align}
with $(\theta_1,\theta_2)\in\Omega_\varepsilon$.
\end{Lem}
\emph{Proof:} See Appendix \ref{app:ecresultfix}. \hfill$\square$

\begin{Rem}
The above theorem covers all the possibilities that symmetric or asymmetric delay constraints on the queues at the source and relay node can be optimal for achieving the maximum effective capacity of the relay system. \textbf{Case I} refers to the case that the maximum throughput can be achieved with symmetric delay constraints at the queues of the source and relay node. \textbf{Case II} represents the case when the statistical delay constraints at the relay can be more stringent, while \textbf{Case III} shows the scenario for more strict delay constraints at the source. Recalling \cite[Theorem 2]{deli-twohop}, we know that as $\varepsilon\to1$, $\theta_1\to0$ and $\theta_2\to0$, and hence
\begin{align}
\lim_{\varepsilon\to1}R_\varepsilon(\varepsilon,D_\tmax)&=\min\left\{\lim_{\theta_1\to0}\frac{J_1(\theta_1)}{\theta_1},\lim_{\theta_1\to0}\frac{J_2(\theta_2)}{\Lambda_{p_1}(\theta_2,\mathcal{Z})}\right\}\\
&=\min\left\{\E\{C_s\},\frac{\E\{C_r\}}{\Pr\{\mathbf{z}\in\mathcal{Z}\}}\right\}.
\end{align}
\end{Rem}

\subsection{Selection Relaying Protocols}\label{sec:relaypolicy}

Considering the expression of the effective capacity and the associated conditions in Theorem \ref{theo:ecresultfix}, we note that finding the optimal relaying protocol in closed-form analytical expressions seems intractable for a general scenario. With this in mind, we consider a simplified case in which the relaying protocol is decided by a function of $\tilde{z}_{sr}$ and $z_{sd}$, and is denoted as $\tilde{z}_{sr}=g(z_{sd})$. The channel state region $\mathcal{Z}$ is given by
\begin{align}\label{eq:Z}
\mathcal{Z}= \{\mathbf{z}:\tilde{z}_{sr}>g(z_{sd})\}.
\end{align}
Also, assume that the decoding strategy at the destination is given by $z_{rd} = f(z_{sd})$, such that
\begin{align}\label{eq:Z0}
\mathcal{Z}_0=\{\mathbf{z}:z_{rd}>f(z_{sd})\}.
\end{align}

%Also, we need the following definition.
%\begin{Def}
%Consider two functions $g(x)$ and $g(x)$ defined in $\mathfrak{R}+$. We have $g \succeq g'$ iff $g(x)\ge g'(x),\forall x\in\mathfrak{R}^+$.
%\end{Def}
\subsubsection{Max Channel Gain (MCG)}
A typical relay policy with buffer-aided relay system is to select the link with stronger channel condition or received SNR \cite{maxmaxrelay}. Then, we have
\begin{align}
g(z_{sd}) = z_{sd}.
\end{align}
We know that $\mathcal{Z}=\{\mathbf{z}: \tilde{z}_{sr}>z_{sd}\}$. With this relaying scheme $\mathcal{Z}$, we can obtain the effective capacity according to Theorem \ref{theo:ecresultfix}.

\subsubsection{Max Delay Exponent (MDE)}
In this part, we propose a relay policy that takes into account the statistical delay constraints as well. Assume that the optimal statistical queueing constraints $(\theta_1,\theta_2)\in\Omega_\varepsilon$ that maximize the effective capacity are given. With this parameter set $(\theta_1,\theta_2)$, we consider the relay strategy that maximizes the statistical delay exponent $J_1(\theta_1)$ at the source, in which case the effective capacity can be potentially improved. Assume that the channel of the link $\bR-\bD$ is independent of the links $\bS-\bR$ and $\bS-\bD$. Combining (\ref{eq:bfJ1}), (\ref{eq:Z}) and (\ref{eq:Z0}), we can express the statistical delay exponent $J_1(\theta_1,g(z_{sd}))$ at the source as
%(\ref{eq:J1gdef}) on top of next page.
%\begin{figure*}
%\begin{small}
\begin{align}\label{eq:J1gdef}
J_1(\theta_1,g(z_{sd}))%\nonumber\\
%&
&= -\log\Bigg(\int_0^\infty \int_{g(z_{sd})}^\infty (1+\tsnr \tilde{z}_{sr})^{-\beta_1} p(z_{sd},\tilde{z}_{sr}) d\tilde{z}_{sr} dz_{sd} \nonumber\\
&\hspace{.5cm}
+ \int_0^\infty \int_{0}^{g(z_{sd})}\int_{f(z_{sd})}^\infty (1+\tsnr z_{sd})^{-\beta_1}p(z_{sd},\tilde{z}_{sr}) p(z_{rd})dz_{rd}d\tilde{z}_{sr}dz_{sd}\nonumber\\
&\hspace{.5cm} +  \int_0^\infty \int_{0}^{g(z_{sd})}\int_0^{f(z_{sd})} \left(1+\frac{\tsnr z_{sd}}{1+\tsnr_r z_{rd}}\right)^{-\beta_1}p(z_{sd},\tilde{z}_{sr}) p(z_{rd})dz_{rd}d\tilde{z}_{sr}dz_{sd}\Bigg)
\end{align}
%\end{small}
%\hrule
%\end{figure*}
Then, the associated relay strategy should be the solution to the following optimization problem
\begin{align}\label{eq:optproblem-J1}
\max_{g} J_1(\theta_1,g(z_{sd})).
\end{align}
We can obtain the relaying strategy specified as below.
\begin{Lem}
Given $(\theta_1,\theta_2)\in\Omega_\varepsilon$, the relay strategy as a function of $(z_{sd},\tilde{z}_{sr})$ that maximizes the statistical delay exponent at the source is given by
\begin{align}\label{eq:J1relay}
g(z_{sd},\theta_1)=\frac{1}{\tsnr}\left(e^{-\frac{1}{\beta_1}\log\E_{z_{rd}}\left\{e^{-\theta_1 C_{sd,\tMAC}}\right\}}-1\right)
\end{align}
where $\beta_1 = \frac{\theta_1TB}{\log2}$, and
%\begin{small}
\begin{align}\label{eq:csmac}
\hspace{-.5cm}C_{sd,\tMAC} = \left\{\begin{array}{ll}TB\log_2(1+\tsnr z_{sd}),& z_{rd} \ge f(z_{sd}),\\
TB\log_2\left(1+\frac{\tsnr z_{sd}}{1+\tsnr_r z_{rd}}\right),& z_{rd} < f(z_{sd}).\end{array}\right.
\end{align}
%\end{small}
\end{Lem}
\emph{Proof:} Define $\hat{g}(z_{sd}) = g(z_{sd},\theta_1)+s\eta(z_{sd})$, where $g(z_{sd},\theta_1)$ is the optimal function, $s$ is any constant, and $\eta(z_{sd})$ represents arbitrary perturbation. Now, a necessary condition that needs to be satisfied for the solution to (\ref{eq:optproblem-J1}) is \cite{physics}
\begin{align}\label{eq:optcond}
\frac{d}{ds}\left(J_1(\theta_1,\hat{g}(z_{sd}))\right)\bigg|_{s=0}=0.
\end{align}
By noting that $\frac{d\hat{g}(z_{sd})}{ds}=\eta(z_{sd})$, and from (\ref{eq:J1gdef}) and (\ref{eq:optcond}), we can
obtain
%derive (\ref{eq:optfirstder}) on top of next page.
%\begin{figure*}
%\begin{small}
\begin{align}\label{eq:optfirstder}
&\int_0^\infty\bigg(-\left(1+\tsnr g(z_{sd},\theta_1)\right)^{-\beta_1}+ \int_{f(z_{sd})}^\infty (1+\tsnr z_{sd})^{-\beta_1} p(z_{rd})dz_{rd} \nonumber\\
&+ \int_0^{f(z_{sd})}\left(1+\frac{\tsnr z_{sd}}{1+\tsnr_r z_{rd}}\right)^{-\beta_1} p(z_{rd})dz_{rd} \bigg)p(z_{sd},g(z_{sd},\theta_1))\eta(z_{sd})dz_{sd} =0
\end{align}
%\end{small}
%\hrule
%\end{figure*}
Since the above equation holds for any $\eta(z_{sd})$, it follows that
%\begin{small}
\begin{align}
&-\left(1+\tsnr g(z_{sd},\theta_1)\right)^{-\beta_1}+ \int_{f(z_{sd})}^\infty (1+\tsnr z_{sd})^{-\beta_1} p(z_{rd})dz_{rd} \nonumber\\
&+ \int_0^{f(z_{sd})}\left(1+\frac{\tsnr z_{sd}}{1+\tsnr_r z_{rd}}\right)^{-\beta_1} p(z_{rd})dz_{rd}=0
\end{align}
%\end{small}
which after rearranging and defining $C_{sd,\tMAC}$ shown in (\ref{eq:csmac}) yields (\ref{eq:J1relay}).\hfill$\square$

%\begin{Rem}
%Obviously, $g(z_{sd},\theta_1)$ is decreasing in $\theta_1$ since $-\frac{1}{\beta_1}\log\E_{z_{rd}}\left\{e^{-\theta_1 C_{s,\tMAC}}\right\}$ is decreasing in $\theta_1$. %So it is surprising that for more stringent delay constraints, the source is more likely to select the relay for transmission.
%\end{Rem}

\begin{Lem}\label{theo:gonJ1}
With the optimal $g(z_{sd},\theta_1)$, we can show the following:
\begin{enumerate}[a)]
\item $J_1(\theta_1,g(z_{sd},\theta_1))$ is increasing in $\theta_1$.
\item $\frac{J_1(\theta_1,g(z_{sd},\theta_1))}{\theta_1}$ is decreasing in $\theta_1$.
\end{enumerate}
\end{Lem}
\emph{Proof:} See Appendix \ref{app:gonJ1}.\hfill$\square$

With the above properties, we know that given $J_1$, there is a unique $\theta_1$ with $g(z_{sd},\theta_1)$ achieving $J_1(\theta_1,g(z_{sd},\theta_1))=J_1$, and as $J_1$ increases, $\theta_1$ increases, and $\frac{J_1(\theta_1,g(z_{sd},\theta_1))}{\theta_1}$ decreases. Therefore, as we iterate over $(\theta_1,\theta_2)\in\Omega_\varepsilon$, we know that the achievable rate of the relay system decreases as $\theta_1$, or equivalently $J_1(\theta_1)$, increases. Then, we can still apply the method in Theorem \ref{theo:ecresultfix} to obtain the maximum effective capacity associated with the proposed relay policy. Instead of fixed relaying policy $\mathcal{Z}$, we now have different $\mathcal{Z}=\{\mathbf{z}:\tilde{z}_{sr}>g(z_{sd},\theta_1)\}$ for different $(J_1(\theta_1),J_2(\theta_2))$.

\begin{Rem}
As $\varepsilon\to1$, i.e., $\theta_1\to0$ and $\theta_2\to0$, we have
\begin{align}\label{eq:gzsd0}
g(z_{sd},0)=\frac{1}{\tsnr}\left(e^{\E_{z_{rd}}\{C_{sd,\tMAC}\}} - 1\right).
\end{align}
This is equivalent to selecting the relay when the instantaneous channel rate of $\bS-\bR$ is larger than the average channel rate of $\bS-\bD$ given $z_{sd}$. Note that $g(z_{sd},0)<z_{sd}$, and hence the source may select the relay for transmission even when the link $\bS-\bD$ is better due to the potential interference from the relay. In this case, according to Theorem \ref{theo:ecresultfix}, we have the achievable throughput given by
\begin{small}
\begin{align}\label{eq:eps1val}
\min\left\{\E\{C_s\},\frac{\E\{C_r\}}{\Pr\{\tilde{z}_{sr}>g(z_{sd},0)\}}\right\}
\end{align}
\end{small}
where the first term is the average service rate of the queue at the source, while the second term is the maximum supported rate considering the buffer at the relay.
\end{Rem}

\begin{Rem}
If $f(z_{sd})=0$, i.e., the destination always decodes the signal from $\bS$ last, we know that $g(z_{sd},\theta_1) = z_{sd}$, i.e., the proposed relay policy reduces to the max channel gain selection.
\end{Rem}

\subsubsection{No-Buffer Relay}

For comparison, we also consider the performance of the relay system without buffer at the relay. We assume buffer at the source node only. Information-theoretical analysis have shown that the maximum rate for the three-node relay network with DF is given by \cite{coopdiver}
%\begin{small}
\begin{align}\label{eq:nobufferrate}
C = \frac{TB}{2}\min\{\log_2(1+2\tsnr z_{sr}),\log_2(1+2\tsnr z_{sd} + 2\tsnr z_{rd})\}.
\end{align}
%\end{small}
The effective capacity associated with arbitrary queueing constraints $\theta$ is
%\begin{small}
\begin{align}
R(\theta) = -\frac{1}{\theta} \log\E\{e^{-\theta C}\}.
\end{align}
%\end{small}
Then, to guarantee the statistical delay constraints $(\varepsilon,D_\tmax)$, we should have from (\ref{eq:sddelay})
%\begin{small}
\begin{align}
\theta R(\theta)\ge -\frac{\log \varepsilon}{D_\tmax}.
\end{align}
%\end{small}
Since $R(\theta)$ is decreasing in $\theta$, there must be one smallest $\theta_\tmin$ such that the above inequality holds with equality. The maximum achievable throughput is then given by $R(\theta_\tmin)$.

\section{Numerical Results}

\begin{figure}
\begin{center}
\includegraphics[width=\figsize\textwidth]{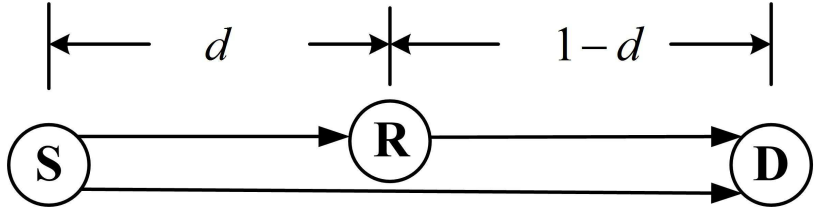}
\caption{The relay model.}\label{fig:systemmodel2}
\end{center}
\end{figure}

We consider the relay model depicted in Fig. \ref{fig:systemmodel2}.
The source, relay, and destination nodes are located on a straight
line. The distance between the source and the destination is
normalized to 1. Let the distance between the source and the relay
node be $d \in(0,1)$. Then, the distance between the relay and
the destination is $1-d$.
We assume the fading distributions for
$\bS-\bD$, $\bS-\bR$ and $\bR-\bD$ links follow independent Rayleigh fading with means $\E\{z_{sd}\} = 1$,
$\E\{z_{sr}\} = 1/d^\alpha$ and $\E\{z_{rd}\} = 1/(1-d)^\alpha$, respectively, where we assume that the path loss $\alpha=4$. We assume $\tsnr=0$ dB, $T=1$ ms, $B=180$ kHz, and $f(z_{sd})=z_{sd}$ in the following numerical results. It has been demonstrated that almost ideal full-duplex relaying can be achieved (1.87$\times$ gain in median throughput compared with 2$\times$ of the ideal case (see \cite[Section 5.3]{fd-2})). Therefore, without loss of generality, we assume ideal full-duplex relay with $\gamma=1$, i.e., $\tilde{z}_{sr} = z_{sr}$. Note that larger $\gamma$ implies stronger residual self-interference at the relay, which decreases the received SNR at the relay, and hence the achievable throughput of the system.

\begin{figure}
\begin{center}
\includegraphics[width=\figsize\textwidth]{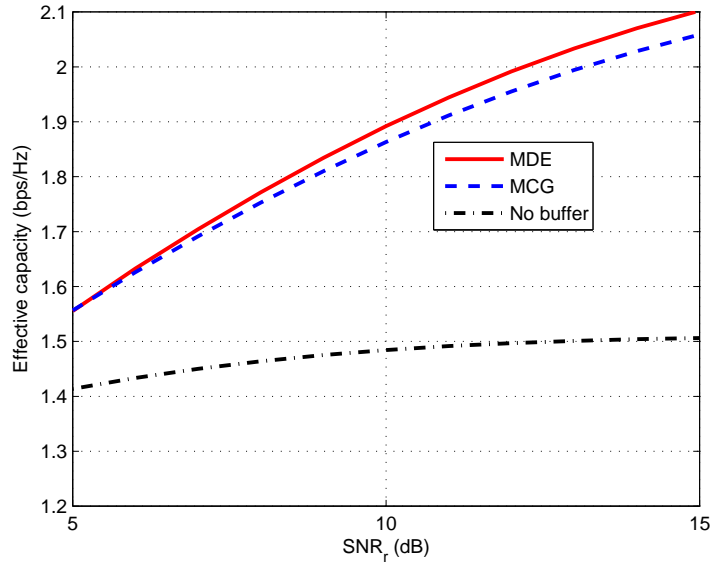}
\caption{Effective capacity as a function of $\tsnr_r$. $\varepsilon=0.05$.%$\tsnr=0$ dB. The solid line represents the relay policy that maximizes the delay constraint at source proposed in this paper, the dashed line stands for the max channel gain relay policy, while the dot-dashed line is the curved achieved with DF policy without buffer at the relay.
}\label{fig:ecinsnr_delaybound=05_dmax=1}
\end{center}
\end{figure}

In Fig. \ref{fig:ecinsnr_delaybound=05_dmax=1}, we plot the effective capacities for the different relay policies as a function of $\tsnr_r$ of the relay node. We assume $d=0.5$, $\varepsilon=0.05$. We can see from the figure that buffering relay can still help improve the system throughput under statistical delay constraints. In addition, the proposed relay policy can achieve better performance than the policy that selects relay with stronger channel condition. As $\tsnr_r$ increases, the achievable throughput without buffer at relay approaches some finite value, since the service rate of the queue at the source is limited by $\frac{TB}{2}\log_2(1+2\tsnr z_{sr})$.

\begin{figure}
\begin{center}
\includegraphics[width=\figsize\textwidth]{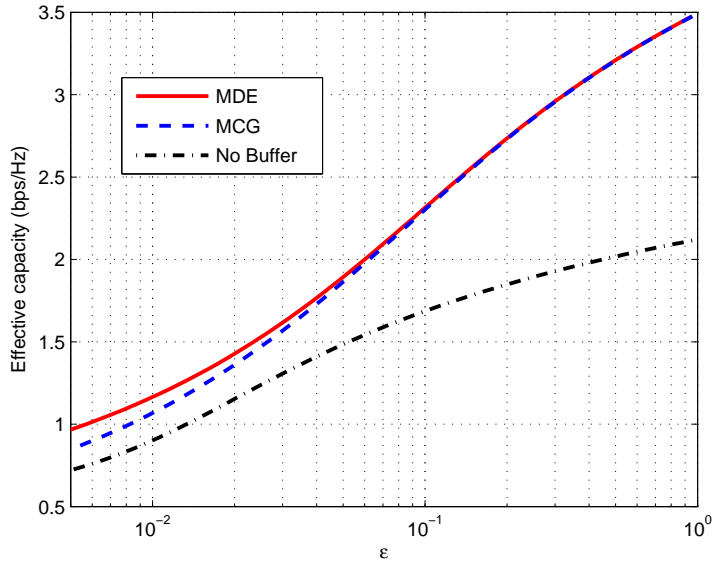}
\caption{Effective capacity as a function of $\varepsilon$. $\tsnr_r=10$ dB. $d = 0.5$.}\label{fig:ecindelaybound}
\end{center}
\end{figure}

In Fig. \ref{fig:ecindelaybound}, we plot the effective capacity as $\varepsilon$ varies for $\tsnr_r = 10$ dB. We assume $d=0.5$. As $\varepsilon\to1$, we know that the effective capacity approaches the value given by (\ref{eq:eps1val}). It is interesting that the proposed relay policy can achieve better performance than the best relay selection policy, and the improvement can be enlarged at smaller $\varepsilon$, i.e., more stringent delay constraints. %Note that eventually as $\varepsilon\to0$, the improvement should vanish since the throughput approaches 0, which is the delay limited capacity of the system.
We can also find that buffering relay helps improve the throughput for a wide range of statistical delay constraints.

\begin{figure}
\begin{center}
\includegraphics[width=\figsize\textwidth]{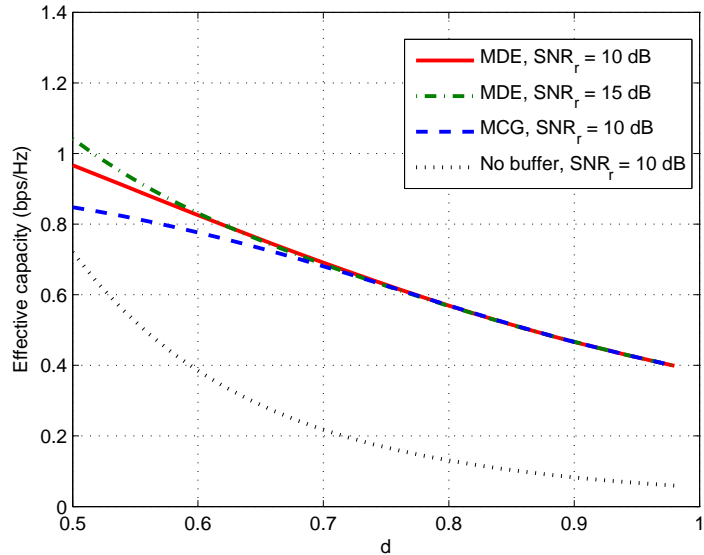}
\caption{Effective capacity as $d$ varies. $\varepsilon=0.005$.}\label{fig:ecind}
\end{center}
\end{figure}

In Fig. \ref{fig:ecind}, we plot the effective capacity as $d$ varies. We assume $\varepsilon=0.005$. We can find that as $d$ increases, i.e., the channel condition between the link $\bS-\bR$ is worse, the effective capacity decreases. The improvement of the proposed relay policy over the best relay selection policy can be larger at smaller $d$, i.e., stronger $\bS-\bR$ link. Also, it is interesting that as $d$ increases, the performance improvement of buffering relay becomes larger. This is mostly because the $\bS-\bR$ now limits the achievable rate of the relay system considering (\ref{eq:nobufferrate}). On the other hand, due to the selection relaying protocol that takes advantage of the link $\bS-\bR$ when its channel is more favorable, significant performance gain can be obtained. Note also that the increase of $\tsnr$ at the relay helps little as $d$ increases, since the buffer at the source is now the bottle-neck of the system. More specifically, as $d$ approaches to 1, the channel between $\bR-\bD$ grows unbounded, i.e., $TB\log_2\left(1+\frac{\tsnr_r z_{rd,\tmin}}{1+\tsnr z_{sd,\tmax}}\right)\ge\max\{TB\log_2(1+\tsnr \tilde{z}_{sr,\tmax}),TB\log_2(1+\tsnr z_{sd,\tmax})\}$. Then, the effective capacity is given by (\ref{eq:ecresultcase2}), where the impact of $\tsnr_r$ on $\theta_{1,0}$ is very small.

\begin{figure}
\begin{center}
\includegraphics[width=\figsize\textwidth]{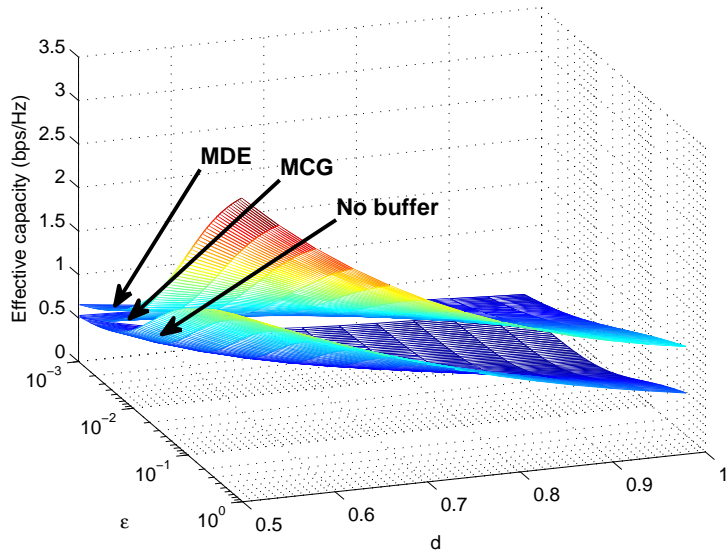}
\caption{Effective capacity as $d$ and $\varepsilon$ vary. $\tsnr_r = 10$ dB. }\label{fig:ecinde}
\end{center}
\end{figure}
In Fig. \ref{fig:ecinde}, we plot the effective capacity as $d$ and $\varepsilon$ vary. We assume $\tsnr_r=10$ dB. We can see from the figure that the improvement provided by the proposed relay policy over the best relay selection policy is more obvious at relatively smaller $\varepsilon$, i.e., more stringent delay constraints. As can be seen from the figure, for all cases considered, the performance of buffering relay systems is better, albeit the improvement vanishes as delay constraints become more stringent.

\begin{figure}
\begin{center}
\includegraphics[width=\figsize\textwidth]{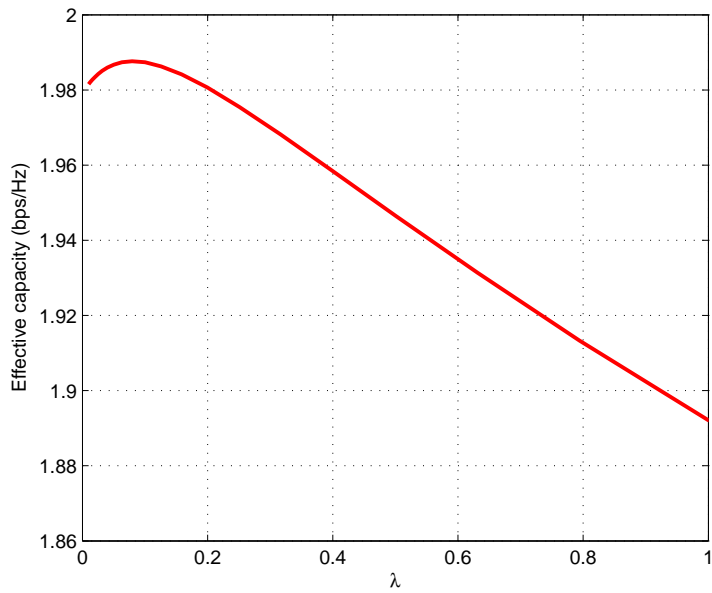}
\caption{Effective capacity vs. $\lambda$. $\tsnr_r = 10$ dB. $d=0.5$. $\varepsilon=0.05$.}\label{fig:ecinlambda}
\end{center}
\end{figure}
Finally, we are interested in the impact of decoding strategy $\mathcal{Z}_0$ in (\ref{eq:Z0}) employed at the destination. Assume $f(z_{sd}) = \lambda z_{sd}$. We plot the effective capacity of the proposed strategy as $\lambda$ varies for $\tsnr_r=10$ dB in Fig. \ref{fig:ecinlambda}. It is interesting that there is an optimal $\lambda$ that maximizes the effective capacity, after which value the effective capacity starts to decrease as $\lambda$ increases. Note that as $\lambda$ increases, $\mathcal{Z}_0$ shrinks. Now, the service rate of $\bS-\bD$ decreases, while the service rate of $\bR-\bD$ increases according to (\ref{eq:rates}) and (\ref{eq:rater}). Hence, as $\mathcal{Z}_0$ shrinks, while the improvement in the link $\bR-\bD$ helps increase the throughput, the deterioration in the link $\bS-\bD$ decreases the throughput. There should be a tradeoff such that the overall throughput can be optimized. Finding the optimal decoding strategy is beyond the topic of this work, and is left for future extension.

\section{Conclusion}
In this paper, we have analyzed the maximum arrival rates that can
be supported by a full-duplex buffer-aided relaying system under statistical delay constraints. We have assumed that both the source and the relay have perfect CSI of the all links, while the destination is not aware of the CSI of source-relay link. We have assumed that the source selects the destination or the relay for data reception based on the selection relaying strategy, which is informed to the destination by the source through an ACK signal such that the destination can perform simultaneous decoding of the received signal.
Given arbitrary selection relay policy, we have expressed the effective bandwidth of the departure process from the queue at the source.
We have obtained the maximum effective capacity under statistical end-to-end delay constraints. In the subsequent
analysis, we have proposed a relay policy that takes into account the statistical delay constraints as well. Through numerical results, we have shown that the delay constraint based relay scheme can achieve better performance than the max channel gain selection policy. Also, we have found that buffering relay can still help improve the throughput in the presence of statistical delay constraints and direct transmission link.

\appendix

%\subsection{Proof of Lemma \ref{lemm:lambdap}}\label{app:lambdap}

\subsection{Proof of Proposition \ref{prop:upperbound}}\label{app:upperbound}
For given $\theta_1$ and $\theta_2$, we can see from (\ref{eq:theta1cond}) that
\begin{align}\label{eq:ecbound1}
R=-\frac{1}{\tilde{\theta}}\log\E\left\{e^{-\tilde{\theta}C_s}\right\}\le-\frac{1}{\theta_1}\log\E\left\{e^{-\theta_1C_s}\right\}.
\end{align}
Note that the above inequality follows immediately from the assumption that $\tilde{\theta}\ge\theta_1$ and $-\frac{\Lambda_{C,1}(-\tilde{\theta})}{\tilde{\theta}}= -\frac{1}{\tilde{\theta}}\log\E\left\{e^{-\tilde{\theta}C_s}\right\}$ is a decreasing function of $\tilde{\theta}$.

Proving another upperbound is a little tricky. Consider the idealistic scenario in which the $\bS-\bD$ and $\bS-\bR$ links are deterministic (i.e., there is no fading) and additionally
can support any constant arrival rate $R>-\frac{1}{\Lambda_{p_1}(\theta_2)}\log\E\left\{e^{-\theta_2 C_{r}}\right\}$. Now, the arriving data to the source can immediately be sent without waiting and consequently there is no need for buffering at the source. Hence, any source queueing constraints can be satisfied. Moreover, assume the source selects relay for transmission with probability $\Pr\{\mathbf{z}\in\mathcal{Z}\}$. With this routing decision, we can see that $\Lambda_{p_1}(\theta_1,\mathcal{Z})$ in (\ref{eq:lmgf-routing}) is unchanged.
If the service rate matches the constant arrival rate, i.e., $C_s = R$, the equation in (\ref{eq:queue1cond}) holds for any $\tilde{\theta}$. That is
\begin{align}
R=-\frac{\Lambda_{C,1}(-\tilde{\theta})}{\tilde{\theta}} =-\frac{1}{\tilde{\theta}}\log\E\left\{e^{-\tilde{\theta}R}\right\} = -\frac{1}{\tilde{\theta}}(-\tilde{\theta} R) = R
\end{align}
where the instantaneous service rate is assumed to be equal to the constant arrival rate $R$. Since no buffering is now required at the source, we can freely impose the most strict QoS constraints and assume $\tilde{\theta}$ to be unbounded as well. Then, with the relay strategy and Lemma \ref{lemm:lambdap}, we have $\Lambda_{p_1}(\hat{\theta},\mathcal{Z})\le\hat{\theta}\le\tilde{\theta}$ for any $\hat{\theta}$. With this, we see from (\ref{eq:queue2cond}) that
\begin{align}\label{eq:ecbound2}
R=-\frac{1}{\Lambda_{p_1}(\hat{\theta},\mathcal{Z})}\log\E\left\{e^{-\hat{\theta}C_{r}}\right\}\le-\frac{1}{\Lambda_{p_1}(\theta_2,\mathcal{Z})}\log\E\left\{e^{-\theta_2 C_{r}}\right\}
\end{align}
where the inequality follows since $-\frac{1}{\Lambda_{p_1}(\theta,\mathcal{Z})}\log\E\left\{e^{-\theta C_{r}}\right\}$ is decreasing in $\theta$ and $\hat{\theta}\ge\theta_2$. Combining the bounds in (\ref{eq:ecbound1}) and (\ref{eq:ecbound2}), we can equivalently write
\begin{align}
R\le\min\left\{-\frac{1}{\theta_1}\log\E\left\{e^{-\theta_1 C_s}\right\}, -\frac{1}{\Lambda_{p_1}(\theta_2,\mathcal{Z})}\log\E\left\{e^{-\theta_2 C_{r}}\right\}\right\}.
\end{align}
\hfill$\square$

\subsection{Proof of Theorem \ref{theo:ecresultfix}}\label{app:ecresultfix}

The idea of this proof follows that in \cite[Appendix C]{deli-twohopend}, except that the effective bandwidth of the arrival process to the queue at the relay is now given by (\ref{eq:relayareb}), and the upperbound on the effective capacity is given by (\ref{eq:upperboundrate}) instead. We here give the sketch of the proof.

We seek to find the optimal $J_1(\theta_1)$ and $J_2(\theta_2)$ on the lower boundary curve specified by Lemma \ref{lemm:J1J2relation}. We first start from the point $J_1(\theta_1)=J_2(\theta_2)=J_{th}(\varepsilon)$. From (\ref{eq:bfJ1}) and (\ref{eq:bfJ2}), we can obtain the associated $\theta_{1,th}$ and $\theta_{2,th}$, which are solutions to
\begin{align}
J_1(\theta_{1,th})=J_{th}(\varepsilon),\,\text{and}\,J_2(\theta_{2,th})=J_{th}(\varepsilon).
\end{align}
Compute $\Lambda_{p_1}(\theta_{2,th},\mathcal{Z})$ from (\ref{eq:routingdef}). Depending on the relationship between $\theta_{1,th}$ and $\Lambda_{p_1}(\theta_{2,th},\mathcal{Z})$, we have three different cases.

\underline{\textbf{Case I}:} Assume $\theta_{1,th}=\Lambda_{p_1}(\theta_{2,th})$. We can show that the effective capacity of the relay system is given by
\begin{align}
R_\varepsilon(\varepsilon,D_\tmax) = \sup_{(\theta_1,\theta_2)\in\Omega}R_E(\theta_1,\theta_2) = R_E(\theta_{1,th},\theta_{2,th}) = \frac{J_{th}(\varepsilon)}{\theta_{1,th}}=\frac{J_{th}(\varepsilon)}{\Lambda_{p_1}(\theta_{2,th},\mathcal{Z})},
\end{align}
considering Proposition \ref{prop:upperbound}.

\underline{\textbf{Case II}:} Assume $\theta_{1,th}>\Lambda_{p_1}(\theta_{2,th},\mathcal{Z})$. In this case, we can relieve the queueing constraints at the source, i.e., decrease $\theta_1$, or $J_1(\theta_1)$ equivalently. Correspondingly, according to Lemma \ref{lemm:J1J2relation}, $J_2(\theta_2)$, and hence $\theta_2$, should increase. We can show that the queue at the relay will not affect the performance as long as $\theta_1$ and $\Lambda_{p_1}(\theta_2,\mathcal{Z})$ satisfies the following inequality given by
\begin{align}\label{eq:ineqcond}
-\frac{1}{\theta_1}\log\E\left\{e^{-\theta_1 C_{s}}\right\}&\le -\frac{1}{\theta_1}\left(\log\E\left\{e^{-\theta_2 C_{r}}\right\} +\log\E\left\{e^{(\Lambda_{p_1}(\theta_2,\mathcal{Z})-\theta_1)C_s}\right\}\right)\nonumber\\
& = \frac{\theta_2}{\theta_1}\left(-\frac{1}{\theta_2}\log\E\left\{e^{-\theta_2 C_{r}}\right\} -\frac{1}{\theta_2}\log\E\left\{e^{(\Lambda_{p_1}(\theta_2,\mathcal{Z})-\theta_1)C_s}\right\}\right).
\end{align}
Note that in the limit as $J_2(\theta_2)\to\infty$, or $\theta_2\to\infty$, we have
\begin{align}
\lim_{\theta_2\to\infty}-\frac{1}{\theta_2}\log\E\left\{e^{-\theta_2 C_{r}}\right\} = \min C_{r} = TB\log_2\left(1+\frac{\tsnr_r z_{rd,\tmin}}{1+\tsnr z_{sd,\tmax}}\right)
\end{align}
the minimum service rate of the queue at relay. Also,
\begin{align}
\lim_{\theta_2\to\infty}\frac{1}{\theta_2}\log\E\left\{e^{(\Lambda_{p_1}(\theta_2,\mathcal{Z})-\theta_1)C_s}\right\}=\max C_{s} = \max\{TB\log_2(1+\tsnr \tilde{z}_{sr,\tmax}), TB\log_2(1+\tsnr z_{sd,\tmax})\}
\end{align}
the maximum service rate of the queue at source. If $TB\log_2\left(1+\frac{\tsnr_r z_{rd,\tmin}}{1+\tsnr z_{sd,\tmax}}\right)>\max\{TB\log_2(1+\tsnr \tilde{z}_{sr,\tmax}), TB\log_2(1+\tsnr z_{sd,\tmax})\}$, we know that there is no congestion at the relay node at all. Then, the queue at the source is the bottle-neck of the relay system, and the effective capacity is
\begin{align}
R_\varepsilon(\varepsilon,D_\tmax) = \frac{J_0}{\theta_{1,0}}
\end{align}
where $J_0$ is defined in (\ref{eq:J1J2relation}), and $\theta_{1,0}$ is the solution to $J_1(\theta_1)=J_0$. Otherwise, we have $(\vvtheta_1,\vvtheta_2)\in\Omega_\varepsilon$ such that $\vvtheta_1$ is the smallest value of $\theta_1$ while (\ref{eq:ineqcond}) can be satisfied with equality at $(\theta_1,\theta_2)$. Now, we can show that the effective capacity is given by
\begin{align}
R_\varepsilon(\varepsilon,D_\tmax) = \sup_{(\theta_1,\theta_2)\in\Omega}R_E(\theta_1,\theta_2) = R_E(\vvtheta_1,\vvtheta_2)=\frac{J_1(\vvtheta_1)}{\vvtheta_1}.
\end{align}
Also, similar to \cite[Lemma 4]{deli-twohopend}, we can show the condition for the uniqueness of $(\vvtheta_1,\vvtheta_2)$.

\underline{\textbf{Case III}:} Assume $\theta_{1,th}<\Lambda_{p_1}(\theta_{2,th},\mathcal{Z})$. In this case, we should relieve the queueing constraints at the relay instead, i.e., decrease $\theta_2$, or $J_2(\theta_2)$ equivalently. So we should have larger $\theta_1$, or $J_1(\theta_1)$, from Lemma \ref{lemm:J1J2relation}. Then, we know that the effective capacity with $\theta_1>\Lambda_{p_1}(\theta_2,\mathcal{Z})$ is given by
\begin{align}
\min\left\{-\frac{1}{\theta_1}\log\E\left\{e^{-\theta_1 C_{s}}\right\}, -\frac{1}{\Lambda_{p_1}(\theta_2,\mathcal{Z})}\log\E\left\{e^{-\theta_2 C_{r}}\right\}\right\}.
\end{align}
Note that in the limit as $J_1(\theta_1)\to\infty$, or $\theta_1\to\infty$, we have
\begin{align}
\lim_{\theta_1\to\infty}-\frac{1}{\theta_1}\log\E\left\{e^{-\theta_1 C_{s}}\right\}&=\min C_s \nonumber\\
&= \min\left\{TB\log_2(1+\tsnr \tilde{z}_{sr,\tmin}),TB\log_2\left(1+\frac{\tsnr z_{sd,\tmin}}{1+\tsnr_r z_{rd,\tmax}}\right)\right\}
\end{align}
the minimum service rate of the queue at the source. Also, as $J_1(\theta_1)\to\infty$, we have $J_2(\theta_2)\to J_0$ from Lemma \ref{lemm:J1J2relation}, and $\theta_2\to\theta_{2,0}$, which is the solution to $J_2(\theta_2)=J_0$. Now
\begin{align}
\lim_{\theta_1\to\infty}-\frac{1}{\Lambda_{p_1}(\theta_2,\mathcal{Z})}\log\E\left\{e^{-\theta_2 C_{r}}\right\}=\frac{J_0}{\Lambda_{p_1}(\theta_{2,0},\mathcal{Z})}.
\end{align}
If $\min\left\{TB\log_2(1+\tsnr \tilde{z}_{sr,\tmin}),TB\log_2\left(1+\frac{\tsnr z_{sd,\tmin}}{1+\tsnr_r z_{rd,\tmax}}\right)\right\} > \frac{J_0}{\Lambda_{p_1}(\theta_{2,0},\mathcal{Z})}$, we know that the queue at the relay is the bottle-neck of the relay system, and the effective capacity is
\begin{align}
R_\varepsilon(\varepsilon,D_\tmax)=\frac{J_0}{\Lambda_{p_1}(\theta_{2,0},\mathcal{Z})}.
\end{align}
Otherwise, we can find a unique pair of $(\uutheta_1,\uutheta_2)\in\Omega_\varepsilon$ such that $\frac{J_1(\uutheta_1)}{\uutheta_1}=\frac{J_2(\uutheta_2)}{\Lambda_{p_1}(\uutheta_2,\mathcal{Z})}$. We can show that the effective capacity is now given by
\begin{align}
R_\varepsilon(\varepsilon,D_\tmax)=\sup_{(\theta_1,\theta_2)\in\Omega}R_E(\theta_1,\theta_2)=R_E(\uutheta_1,\uutheta_2)=\frac{J_2(\uutheta_2)}{\Lambda_{p_1}(\uutheta_2,\mathcal{Z})}.
\end{align}

The details of the derivation for the above claims are similar to the proof in \cite[Appendix C]{deli-twohopend}, and are omitted here. Interested readers are encouraged to find more details in \cite[Appendix C]{deli-twohopend}.
\hfill$\square$

\subsection{Proof of Theorem \ref{theo:gonJ1}}\label{app:gonJ1}
\begin{enumerate}[a)]
\item This can be easily verified since for $\theta_1'>\theta_1$, we have
\begin{align} \hspace{-.6cm}J_1(\theta_1',g(z_{sd},\theta_1'))>J_1(\theta_1',g(z_{sd},\theta_1))>J_1(\theta_1,g(z_{sd},\theta_1))
\end{align}
where the first inequality holds due to the definition of $g(z_{sd},\theta_1')$, and the second inequality holds since given relay policy $g$, $J_1(\theta_1,g)$ is an increasing function of $\theta_1$.

\item Taking the derivative of $J_1(\theta_1,g(z_{sd},\theta_1))$ with respect to $\theta_1$, we have
\begin{align}
&\frac{d}{d\theta_1}J_1(\theta_1,g(z_{sd},\theta_1))\nonumber\\
 &= \frac{1}{\phi_1}\Bigg(\int_0^\infty \int_{g(z_{sd},\theta_1)}^\infty (1+\tsnr \tilde{z}_{sr})^{-\beta_1} TB\log_2(1+\tsnr \tilde{z}_{sr}) p(z_{sd},\tilde{z}_{sr}) d\tilde{z}_{sr} dz_{sd} \nonumber\\
&
+ \int_0^\infty \int_{0}^{g(z_{sd},\theta_1)}\int_{f(z_{sd})}^\infty (1+\tsnr z_{sd})^{-\beta_1}TB\log_2(1+\tsnr z_{sd})p(z_{sd},\tilde{z}_{sr}) p(z_{rd})dz_{rd}d\tilde{z}_{sr}dz_{sd}\nonumber\\
& +  \int_0^\infty \int_{0}^{g(z_{sd},\theta_1)}\int_0^{f(z_{sd})} \left(1+\frac{\tsnr z_{sd}}{1+\tsnr_r z_{rd}}\right)^{-\beta_1}TB\log_2\left(1+\frac{\tsnr z_{sd}}{1+\tsnr_r z_{rd}}\right)p(z_{sd},\tilde{z}_{sr}) p(z_{rd})dz_{rd}d\tilde{z}_{sr}dz_{sd}\nonumber\\
& -\int_0^\infty\bigg(-\left(1+\tsnr g(z_{sd},\theta_1)\right)^{-\beta_1}+ \int_{f(z_{sd})}^\infty (1+\tsnr z_{sd})^{-\beta_1} p(z_{rd})dz_{rd} \nonumber\\
&+ \int_0^{f(z_{sd})}\left(1+\frac{\tsnr z_{sd}}{1+\tsnr_r z_{rd}}\right)^{-\beta_1} p(z_{rd})dz_{rd} \bigg)p(z_{sd},g(z_{sd},\theta_1))\dot{g}(z_{sd},\theta_1)dz_{sd}\Bigg)\label{eq:J1firstderv-proof0}\\
& = \frac{1}{\phi_1}\Bigg(\int_0^\infty \int_{g(z_{sd},\theta_1)}^\infty (1+\tsnr \tilde{z}_{sr})^{-\beta_1} TB\log_2(1+\tsnr \tilde{z}_{sr}) p(z_{sd},\tilde{z}_{sr}) d\tilde{z}_{sr} dz_{sd} \nonumber\\
&
+ \int_0^\infty \int_{0}^{g(z_{sd},\theta_1)}\int_{f(z_{sd})}^\infty (1+\tsnr z_{sd})^{-\beta_1}TB\log_2(1+\tsnr z_{sd})p(z_{sd},\tilde{z}_{sr}) p(z_{rd})dz_{rd}d\tilde{z}_{sr}dz_{sd}\nonumber\\
& +  \int_0^\infty \int_{0}^{g(z_{sd},\theta_1)}\int_0^{f(z_{sd})} \left(1+\frac{\tsnr z_{sd}}{1+\tsnr_r z_{rd}}\right)^{-\beta_1}TB\log_2\left(1+\frac{\tsnr z_{sd}}{1+\tsnr_r z_{rd}}\right)p(z_{sd},\tilde{z}_{sr}) p(z_{rd})dz_{rd}d\tilde{z}_{sr}dz_{sd}\Bigg)\label{eq:J1firstderv-proof1}
\end{align}
where Leibniz's rule is incorporated to obtain (\ref{eq:J1firstderv-proof0}), $\dot{g}(z_{sd},\theta_1)$ is the first derivative of $g$ with respect to $\theta_1$,
\begin{align}
\phi_1& = \int_0^\infty \int_{g(z_{sd},\theta_1)}^\infty (1+\tsnr \tilde{z}_{sr})^{-\beta_1}p(z_{sd},\tilde{z}_{sr}) d\tilde{z}_{sr} dz_{sd} \nonumber\\
&\hspace{.5cm}
+ \int_0^\infty \int_{0}^{g(z_{sd},\theta_1)}\int_{f(z_{sd})}^\infty (1+\tsnr z_{sd})^{-\beta_1}p(z_{sd},\tilde{z}_{sr}) p(z_{rd})dz_{rd}d\tilde{z}_{sr}dz_{sd}\nonumber\\
&\hspace{.5cm} +  \int_0^\infty \int_{0}^{g(z_{sd},\theta_1)}\int_0^{f(z_{sd})} \left(1+\frac{\tsnr z_{sd}}{1+\tsnr_r z_{rd}}\right)^{-\beta_1}p(z_{sd},\tilde{z}_{sr}) p(z_{rd})dz_{rd}d\tilde{z}_{sr}dz_{sd}
\end{align}
and (\ref{eq:optfirstder}) is substituted into (\ref{eq:J1firstderv-proof0}) to obtain (\ref{eq:J1firstderv-proof1}). It is interesting that (\ref{eq:J1firstderv-proof1}) is the same to the partial derivative of $J_1(\theta_1,g)$ with respect to $\theta_1$ while viewing $g=g(z_{sd},\theta_1)$ as a separate variable, i.e.,
\begin{align}\label{eq:dersubs}
\frac{d}{d\theta_1}J_1(\theta_1,g(z_{sd},\theta_1)) = \frac{\partial}{\partial\theta_1}J_1(\theta_1,g)\big|_{g=g(z_{sd},\theta_1)}.
\end{align}
Note that given $g$, $\frac{J_1(\theta_1,g)}{\theta_1}$ is a decreasing function of $\theta_1$, i.e.,
\begin{align}
\frac{d }{d \theta_1}\left(\frac{J_1(\theta_1,g)}{\theta_1}\right)=\frac{\theta_1 \frac{d}{d\theta_1}J_1(\theta_1,g) -J_1(\theta_1,g)}{\theta_1^2}\le0.
\end{align}
Then, combining (\ref{eq:dersubs}), we can show that
\begin{align}
\frac{d}{d\theta_1}\left(\frac{J_1(\theta_1,g(z_{sd},\theta_1))}{\theta_1}\right)&=\frac{\theta_1\frac{d}{d\theta_1}J_1(\theta_1,g(z_{sd},\theta_1)) -J_1(\theta_1,g(z_{sd},\theta_1))}{\theta_1^2} \nonumber\\
&= \frac{\theta_1\frac{\partial}{\partial\theta_1}J_1(\theta_1,g)\big|_{g=g(z_{sd},\theta_1)}  -J_1(\theta_1,g(z_{sd},\theta_1))}{\theta_1^2}\nonumber\\
& =\frac{d }{d \theta_1}\left(\frac{J_1(\theta_1,g)}{\theta_1}\right)\bigg|_{g=g(z_{sd},\theta_1)}\le0.
\end{align}
i.e., the first derivative of $\frac{J_1(\theta_1,g(z_{sd},\theta_1))}{\theta_1}$ with respect to $\theta_1$ is less than zero as well. This proves the result in the theorem.
\end{enumerate}
\hfill$\square$
%\end{spacing}

\end{document}